\documentclass[conference]{IEEEtran}
\IEEEoverridecommandlockouts

\usepackage{cite}
\usepackage{amsmath,amssymb,amsfonts}
\usepackage{algorithmic}
\usepackage{graphicx}
\usepackage{textcomp}
\usepackage{xcolor}
\usepackage{bm}
\usepackage{adjustbox}
\usepackage{algorithm}
\usepackage{subfigure}

\def\BibTeX{{\rm B\kern-.05em{\sc i\kern-.025em b}\kern-.08em
    T\kern-.1667em\lower.7ex\hbox{E}\kern-.125emX}}

\begin{document}

\title{RF-Based Simultaneous Localization and Source Seeking for Multi-Robot Systems\\} 

\author{Ke Xu\textsuperscript{\textasteriskcentered}, Rui Zhang\textsuperscript{\textasteriskcentered\textdagger}, He (Henry) Chen\textsuperscript{\textasteriskcentered\textdagger}\\
\textsuperscript{\textasteriskcentered}Department of Information Engineering, The Chinese University of Hong Kong\\
\textsuperscript{\textdagger}Shun Hing Institute of Advanced Engineering, The Chinese University of Hong Kong\\
\{xk020, ruizhang, he.chen\}@ie.cuhk.edu.hk\thanks{This research was supported in part by project \#MMT 79/22 of the Shun Hing Institute of Advanced Engineering, The Chinese University of Hong Kong.}
}

\maketitle

\begin{abstract}
 This paper considers a radio-frequency (RF)-based simultaneous localization and source-seeking (SLASS) problem in multi-robot systems, where multiple robots jointly localize themselves and an RF source using distance-only measurements extracted from RF signals and then control themselves to approach the source. We design a Rao-Blackwellized particle filter-based algorithm to realize the joint localization of the robots and the source. We also devise an information-theoretic control policy for the robots to approach the source. In our control policy, we maximize the predicted mutual information between the source position and the distance measurements, conditioned on the robot positions, to incorporate the robot localization uncertainties. A projected gradient ascent method is adopted to solve the mutual information maximization problem. Simulation results show that the proposed SLASS framework outperforms two benchmarks in terms of the root mean square error (RMSE) of the estimated source position and the decline of the distances between the robots and the source, indicating more effective approaching of the robots to the source.
\end{abstract}

\begin{IEEEkeywords}
source seeking, cooperative localization, Rao-Blackwellized particle filter, mutual information, multi-robot systems
\end{IEEEkeywords}

\section{Introduction}

Recently, robot systems have been increasingly popular and used to perform a large number of tasks in various scenarios \cite{michael2014collaborative,englot2013three,seeni2010robot}. Compared to a single robot, multi-robot systems are more powerful because the robots can simultaneously make observations of the environment, share the gathered information with each other, and complete the tasks in a more efficient manner. In addition, the inter-robot cooperation and observation diversity make the multi-robot system more robust to network failure. Therefore, multi-robot systems are particularly suitable for complex tasks  \cite{yan2013survey}.

Radio-frequency (RF) source seeking is an important application of multi-robot systems \cite{zou2015particle,charrow2014cooperative,atanasov2015distributed,shaikhanov2022falcon}. More specifically, a team of robots can be used to cooperatively seek an RF source in the areas inaccessible to humans, such as an extraterrestrial surface. The completion of such tasks consists of two steps: (1) determine the source position, and (2) design an appropriate control policy for the robots to approach the source. 
Several source-seeking policies have been proposed in the literature. In \cite{zou2015particle}, a particle swarm optimization (PSO) technique was applied to design the control policy, and the performance of several variations of PSO was compared. In \cite{charrow2014cooperative}, the authors used the particle filter to estimate the source position and designed an information-theoretic control policy by maximizing the mutual information between the source position and the measurements in the next control cycle. In \cite{atanasov2015distributed}, a stochastic gradient ascent method was developed to approximate the exact gradient of the mutual information. A networked drone-based RF source-seeking method was investigated in \cite{shaikhanov2022falcon}, and the results showed that the localization accuracy can be improved if the drones spread out. However, spreading will result in additional flying distance, which is not favored from the perspective of source approaching. To handle such a tradeoff, the authors proposed a unified objective function in their formulated problem to balance the localization accuracy and approaching efficiency, offering a flexible strategy for flight planning.

In all the source-seeking schemes mentioned above, the positions of the robots are assumed to be known, which are obtained using other localization means, such as the global navigation and satellite system (GNSS) or laser scanners. As such, only the position of the source needs to be estimated in each control cycle. However, these localization means may not work in some tasks. For example, laser scanners become less efficient in feature-sparse environments, and GNSS is no longer available for the tasks of space exploration. In these tasks, RF-based localization approaches represent a promising solution to realize robot self-localization \cite{zhang2019distributed,di2017calibration,zhu2021decentralized}, where robots can exchange RF signals to cooperatively localize themselves. With this in mind, it is natural to come up with a trivial two-stage strategy for these tasks: 
the robots first perform RF-based cooperative localization and then execute one of the existing source-seeking algorithms that treat the estimated robot positions as the known ones to further localize the source and obtain the control inputs for all robots. However, one can quickly realize that such a trivial solution overlooks the mutual benefits between the robot self-localization and the source localization that are both based on RF signals. It is apparent that a better estimation of the robot positions will lead to a better estimation of the source position. Conversely, a better estimation of the source position will also benefit the estimation of the robot positions because the source can serve as a fixed reference node for the robots. In this context, we are motivated to develop a new simultaneous localization and source-seeking (SLASS) algorithm to harness the mutual benefits between the robot localization and the source localization. To our best knowledge, such a problem has not been thoroughly investigated in the open literature.

In this paper, we consider a SLASS problem with the joint localization of the source and robots using distance-only measurements, which can be extracted from the time-of-arrival (ToA) parameters of RF signals. In the considered problem, the robot positions are no longer known \textit{a priori} due to the control error and localization uncertainty, making the problem more challenging than existing ones. In particular, the localization of both robots and source needs to be performed in a probabilistic manner, the objective of which is to compute the posterior distributions of the robot and source positions. Based on the obtained posterior distributions, we then adopt a new conditional mutual information as the objective function to design the control policy. We provide a centralized solution to the considered SLASS problem by developing new algorithms for both cooperative localization and information-theoretic control. For the localization part, we apply a Rao-Blackwellized particle filter \cite{murphy2001rao} to approximate the continuous distributions as discrete ones and compute the marginalized posterior probabilities of the involved unknown variables. For the control part, we use a projected gradient ascent method to navigate the robots to the directions with the largest mutual information. Simulation results are provided to demonstrate the merits of the proposed algorithms.

\section{System Model and Problem Formulation}\label{model_section}

We consider a system consisting of $K$ robots and a remote and stationary RF source. The position of robot $k$ at the $n$-th control cycle is denoted by $\mathbf{x}_{n,k}=[x_{n,k},y_{n,k}]^T$, and the source is located at $\mathbf{x}_{n,0}=[x_{0},y_{0}]^T$ with $n=1,\cdots,N$. The task of the robots is to control themselves to approach the source. The positions of the robots and the source are both unknown, and thus we will use a joint probabilistic model to simultaneously estimate their positions. Assume the robots can receive reference RF signals to estimate the distances between each other and between the source and themselves. In this paper, we use the following distance measurement model \cite{charrow2014cooperative}
\begin{equation}\label{measurement model}
	\begin{aligned}
		&p(z_{n,k_1,k_2}|\mathbf{x}_{n,k_1},\mathbf{x}_{n,k_2})\\
		&=\mathcal{N}(z_{n,k_1,k_2};\alpha_0+r_{n,k_1,k_2}\alpha,r_{n,k_1,k_2}\sigma_z^2),
	\end{aligned}
\end{equation}
where $z_{n,k_1,k_2}$ represents the distance measurement obtained from the RF signal received by robot $k_2~(k_2=1,\cdots,K)$ and transmitted from robot $k_1$ ($k_1=1,\cdots,K$) or the source $(k_1=0)$. 
$\mathcal{N}(z;\mu,\sigma^2)$ is the probability density function of a Gaussian random variable $z$ with mean $\mu$ and variance $\sigma^2$. $r_{n,k_1,k_2}=\lVert\mathbf{x}_{n,k_1}-\mathbf{x}_{n,k_2}\rVert_2$ is the distance between node $k_1$ and node $k_2$. $\alpha_0$, $\alpha$, and $\sigma_z^2$ are parameters related to the physical environment.\footnote{These parameters can be estimated using some labeled data \cite{charrow2014cooperative}. }

Based on (\ref{measurement model}), the joint likelihood function at the $n$-th control cycle can be written as
\begin{equation}\label{likelihood}
	\begin{aligned}
		&p(\mathbf{z}_n|\mathbf{x}_n)
		= \prod_{k=1}^Kp(z_{n,0,k}|\mathbf{x}_{n,0},\mathbf{x}_{n,k})\\
		&\quad\quad\quad\quad\quad\times\prod_{1\leq k_1< k_2\leq K}p(z_{n,k_1,k_2}|\mathbf{x}_{n,k_1},\mathbf{x}_{n,k_2}),
	\end{aligned}
\end{equation}
where $\mathbf{z}_n=[z_{n,0,1},\cdots,z_{n,0,K},z_{n,1,2},\cdots,z_{n,K-1,K}]^T$ is the collection of all the measurements between any two nodes, and $\mathbf{x}_n=[\mathbf{x}_{n,0}^T,\mathbf{x}_{n,1}^T,\cdots,\mathbf{x}_{n,K}^T]^T$ is the collection of all the node positions. In the above equation, we assume that the measurements are mutually independent given the positions of the source and the robots.

In each control cycle, the robots need to generate control inputs to decide where they should move in the next step based on the distance measurements. The relationship of the robot positions in two consecutive control cycles is described as
\begin{equation}\label{transition relation}
	\mathbf{x}_{n,k}=\mathbf{x}_{n-1,k}+\mathbf{c}_{n-1,k}+\mathbf{n}_{n-1,k},
\end{equation}
where $\mathbf{c}_{n-1,k}$ is the control input of robot $k$ at the $(n-1)$-th control cycle, and $\mathbf{n}_{n-1,k}$ is the control error following a Gaussian distribution with zero mean and variance $\sigma_c^2$. Hence, the transition distribution of the robot position is given by
\begin{equation}\label{transition probability robot}
	p(\mathbf{x}_{n,k}|\mathbf{x}_{n-1,k})=\mathcal{N}(\mathbf{x}_{n,k};\mathbf{x}_{n-1,k}+\mathbf{c}_{n-1,k},\sigma_c^2\mathbf{I}),
\end{equation}
where $\mathbf{I}$ is an identity matrix. In addition, since the source remains stationary during the whole seeking process, its transition distribution of the position can be written as
\begin{equation}\label{transition probability source}
	p(\mathbf{x}_{n,0}|\mathbf{x}_{n-1,0})=\delta(\mathbf{x}_{n,0}-\mathbf{x}_{n-1,0}),
\end{equation}
where $\delta(\cdot)$ is a Dirac impulse function. 

At the initialization stage, $p(\mathbf{x}_{1,0})\triangleq p(\mathbf{x}_{1,0}|\mathbf{x}_{0,0})$ is assigned to be a uniform distribution in the explored area, and  $p(\mathbf{x}_{1,k})\triangleq p(\mathbf{x}_{1,k}|\mathbf{x}_{0,k}) =\delta(\mathbf{x}_{1,k}-\bar{\mathbf{x}}_{1,k})$ for $k=1,\cdots,K$, where $\bar{\mathbf{x}}_{1,k}$ is the true starting position of robot $k$. It should be noted that although the robot positions are assumed to be known at $n=1$, it is inadequate to use the initial positions and the control inputs to infer the robot positions when $n\geq2$. This is because the control error $\mathbf{n}_{n,k}$ can accumulate over time and the estimates of robot positions will become highly inaccurate if the error is simply ignored.

Combining the likelihood function in (\ref{likelihood}) and the transition prior distributions in (\ref{transition probability robot}) and (\ref{transition probability source}), the joint posterior distribution of the source and robot positions can be computed as
\begin{equation}\label{joint distribution}
	\begin{aligned}
		p(\mathbf{x}_{1:N}|\mathbf{z}_{1:N})&\propto p(\mathbf{x}_{1:N})p(\mathbf{z}_{1:N}|\mathbf{x}_{1:N})\\
		&\propto \prod_{n=1}^Np(\mathbf{x}_{n}|\mathbf{x}_{n-1})p(\mathbf{z}_n|\mathbf{x}_n),
	\end{aligned}
\end{equation}
where $\mathbf{x}_{1:N}=[\mathbf{x}_1^T,\cdots,\mathbf{x}_N^T]^T$, $\mathbf{z}_{1:N}=[\mathbf{z}_1^T,\cdots,\mathbf{z}_N^T]^T$,  $p(\mathbf{x}_{n}|\mathbf{x}_{n-1})=p(\mathbf{x}_{n,0}|\mathbf{x}_{n-1,0})\prod_{k=1}^Kp(\mathbf{x}_{n,k}|\mathbf{x}_{n-1,k})$, and $p(\mathbf{z}_n|\mathbf{x}_n)$ is given in (\ref{likelihood}).
The objective of the SLASS algorithm is to obtain an estimate of the source and robot positions from the posterior distribution, and accordingly design a policy to control the robots to move to their next positions.

\section{Proposed SLASS Algorithm}

In this section, we propose a centralized SLASS algorithm that simultaneously incorporates the position uncertainty of the source and the robots. First, a Rao-Blackwellized particle filter is used to jointly estimate the posterior probabilities of the source and robot positions. Based on the estimated posterior probabilities, a control policy is designed for the robots by maximizing the predicted mutual information between the source position and the distance measurements, conditioned on the robot positions. 
In practice, one of the $K$ robots should act as a central robot which has high computing power. The central robot gathers the measurements from all the robots, executes the SLASS algorithm, and sends the control inputs back to other robots to navigate them. Nevertheless, the detailed implementation is beyond the scope of this paper.

\subsection{Joint Localization of the Source and Robots}\label{particle formulate}

In this part, we elaborate on the joint localization algorithm for the source and robots based on the established probabilistic model in Section \ref{model_section}. In practice, it is intractable to directly compute the posterior distribution from (\ref{joint distribution}) because the source position and robot positions are coupled with each other. To address this challenge, we refer to \cite{murphy2001rao} and develop a Rao-Blackwellized particle filtering algorithm to perform the localization. To this end, we factorize the posterior distribution in (\ref{joint distribution}) as
\begin{equation}\label{RB factorization}
	p(\mathbf{x}_{1:N}|\mathbf{z}_{1:N})=p(\mathbf{x}_{1:N,1:K}|\mathbf{z}_{1:N})p(\mathbf{x}_{1:N,0}|\mathbf{x}_{1:N,1:K},\mathbf{z}_{1:N}),
\end{equation}
where $\mathbf{x}_{1:N,1:K}=[\mathbf{x}_{1,1:K}^T,\cdots,\mathbf{x}_{N,1:K}^T]^T$ and $\mathbf{x}_{n,1:K}=[\mathbf{x}_{n,1}^T,\cdots,\mathbf{x}_{n,K}^T]^T$. Based on this factorization, the posterior distribution of the robot positions can be approximated by a set of discrete particles, i.e.,
\begin{equation}\label{particle robot}
	p(\mathbf{x}_{1:N,1:K}|\mathbf{z}_{1:N})\approx \sum_{i=1}^{M_r} w_N^{(i)}\delta(\mathbf{x}_{1:N,1:K}-\mathbf{x}_{1:N,1:K}^{(i)}),
\end{equation}
where $M_r$ is the number of robot particles, $\mathbf{x}_{1:N,1:K}^{(i)}$ is the $i$-th particle of $\mathbf{x}_{1:N,1:K}$ sampled from an importance density function, which is typically chosen to be the density function of the position transition model, and $w_N^{(i)}$ is the weight of the $i$-th robot particle. Given the robot particle $\mathbf{x}_{1:N,1:K}^{(i)}$, the conditional posterior distribution of the source position can be approximated by
\begin{equation}\label{source robot}
	p(\mathbf{x}_{1:N,0}|\mathbf{x}_{1:N,1:K}^{(i)},\mathbf{z}_{1:N})\approx \sum_{j=1}^{M_s} w_N^{(i,j)}\delta(\mathbf{x}_{1:N,0}-\mathbf{x}_{1:N,0}^{(i,j)}),
\end{equation}
where $M_s$ is the number of source particles subordinate to one robot particle, $\mathbf{x}_{1:N,0}^{(i,j)}$ is the $j$-th particle of $\mathbf{x}_{1:N,0}$ subordinate to robot particle $\mathbf{x}_{1:N,1:K}^{(i)}$ sampled from its importance density function, and $w_N^{(i,j)}$ is the corresponding particle weight. In each control cycle, the weights of the robot and source particles are updated recursively as $w_N^{(i)}\propto w_{N-1}^{(i)}\sum_{j=1}^{M_s} w_{N-1}^{(i,j)}p(\mathbf{z}_N|\mathbf{x}_{N,1:K}^{(i)},\mathbf{x}_{N,0}^{(i,j)})$ and $w_N^{(i,j)}\propto w_{N-1}^{(i,j)}p(\mathbf{z}_N|\mathbf{x}_{N,1:K}^{(i)},\mathbf{x}_{N,0}^{(i,j)})$, respectively. The detailed derivation of the weight update equations is omitted due to space limitation. After updating the particle weights,  resampling should be performed to avoid the degeneracy phenomenon~\cite{arulampalam2002tutorial}.

\subsection{Information-Theoretic Control Policy}\label{inforcontrol}

In this part, we design the control policy by maximizing the predicted mutual information between the source position and the distance measurements, conditioned on the robot positions. We adopt the mutual information as the objective function, as it can effectively measure the reduction in the source position uncertainty when the distance measurements are available in the next control cycle. 

Using the particle representation of the robot positions, the predicted mutual information between the source position and the measurements conditioned on the robot positions is expressed as\footnote{We use the index $N$ instead of $n$ in this part, since we are focusing on the latest control cycle.}
\begin{equation}\label{mutualparticle}
\begin{aligned}	&I(\mathbf{x}_{N+1,0};\mathbf{z}_{N+1,0}|\mathbf{x}_{N+1,1:K})\\
&\approx\sum_{i=1}^{M_r}w_{N+1}^{(i)}I(\mathbf{x}_{N+1,0};\mathbf{z}_{N+1,0}|\mathbf{x}_{N+1,1:K}=\mathbf{x}_{N+1,1:K}^{(i)}),
\end{aligned}
\end{equation}
where $\mathbf{z}_{N+1,0}=[z_{N+1,0,1},\cdots,z_{N+1,0,K}]^T$. Specifically, we assume $w_{N+1}^{(i)}=w_{N}^{(i)}$ and $w_{N+1}^{(i,j)}=w_{N}^{(i,j)}$, and then use the following predicted likelihood function to compute the mutual information in the next control cycle:
\begin{equation}\label{predicted likelihood}
	\begin{aligned}
		&\tilde{p}(\mathbf{z}_{N+1,0}|\mathbf{x}_{N+1,0}^{(i,j)},\mathbf{x}_{N+1,1:K}^{(i)})\\
		&=p(\mathbf{z}_{N,0}|\mathbf{x}_{N,0}^{(i,j)},\mathbf{x}_{N,1:K}^{(i)}+\mathbf{c}_{N,1:K})\bigg|_{\mathbf{z}_{N+1,0}=\mathbf{z}_{N,0}},
	\end{aligned}
\end{equation}
where $\mathbf{c}_{N,1:K}=[\mathbf{c}_{N,1}^T,\cdots,\mathbf{c}_{N,K}^T]^T$. The derivation of the detailed expression of (\ref{mutualparticle}) is omitted due to space limitation. The expression can be derived by adapting the method used in \cite{charrow2014cooperative}.

The control input can be generated by solving the following optimization problem
\begin{equation}\label{max MI}
\begin{aligned}
&\max_{\{\mathbf{c}_{N,k}\}_{k=1}^K}I(\mathbf{x}_{N+1,0};\mathbf{z}_{N+1,0}|\mathbf{x}_{N+1,1:K})  \\
&~~~{\rm{ s}}{\rm{.t}}{\rm{. }} ~ \lVert\mathbf{x}_{N+1,k_1}- \mathbf{x}_{N+1,k_2}\rVert_2\geq d_{\min},\\
&~~~~~~~~\forall k_1,k_2=1,\cdots,K,~k_1\neq k_2,\\
\end{aligned}
\end{equation}
where the constraints are imposed to avoid collisions among the robots. Inspired by \cite{zhang2020self}, we apply a projected gradient ascent method to solve (\ref{max MI}). By maximizing the mutual information, the robots can gradually approach and eventually arrive at the source position. The whole SLASS algorithm is formally presented in Algorithm \ref{A1}.

\begin{figure}
	\label{algorithm111}
	\renewcommand{\algorithmicrequire}{\textbf{Input:}}
	\renewcommand{\algorithmicensure}{\textbf{Output:}}
	\begin{algorithm}[H]
		\begin{small}
			\caption{SLASS at Control Cycle $N$}
			\begin{algorithmic}[1]\label{A1}
				\REQUIRE Robot particles $\{\mathbf{x}_{N,1:K}^{(i)}\}_{i=1}^{M_r}$ and source particles $\{\mathbf{x}_{N,0}^{(i,j)}\}_{i=1,j=1}^{M_r,M_s}$.
				\ENSURE Control inputs of the robots $\{\mathbf{c}_{N,k}\}_{k=1}^K$.
				\STATE The robots receive distance measurements $\mathbf{z}_N$.
				\STATE Compute the weights of source particles $\{w_N^{(i,j)}\}_{i=1,j=1}^{M_r,M_s}$.
				\STATE Compute the weights of robot particles $\{w_N^{(i)}\}_{i=1}^{M_r}$.
				\STATE Normalize the weights and resample the particles.
				\STATE Compute the control inputs of the robots by solving (\ref{max MI}).
				\STATE The robots move to their new positions according to the control inputs.
				\STATE Draw new source particles $\{\mathbf{x}_{N+1,0}^{(i,j)}\}_{i=1,j=1}^{M_r,M_s}$ from $\{p(\mathbf{x}_{N+1,0}|\mathbf{x}_{N,1:K}^{(i)},\mathbf{x}_{N,0}^{(i,j)})\}_{i=1,j=1}^{M_r,M_s}$.
				\STATE Draw new robot particles $\{\mathbf{x}_{N+1,1:K}^{(i)}\}_{i=1}^{M_r}$ from $\{p(\mathbf{x}_{N+1,1:K}|\mathbf{x}_{N,1:K}^{(i)})\}_{i=1}^{M_r}$.
			\end{algorithmic}
		\end{small}
	\end{algorithm}
	\vspace{-3em}
\end{figure}

\section{Simulations}\label{simulation section}
In this section, we evaluate the performance of the proposed SLASS algorithm and show its superiority over the benchmarks through simulations. In the following, we use the meter as the unit of distance-related parameters. A stationary source is located at $\mathbf{x}_{n,0}=[100,100]^T$ for all $n$. The number of robots $K$ varies from $1$ to $3$. The initial positions of the three robots are $\mathbf{x}_{1,1}=[0,0]^T$, $\mathbf{x}_{1,2}=[5,0]^T$ and $\mathbf{x}_{1,3}=[0,5]^T$. As mentioned in Section \ref{model_section}, 
the source particles are initialized to follow a uniform distribution in the explored area $\{[x,y]^T|0\leq x\leq150,0\leq y\leq150\}$, and the initial robot particles are given by $\mathbf{x}_{1,k}^{(i)}=\bar{\mathbf{x}}_{1,k}$ for all $i$ and $k$. The parameters of distance measurements are $\alpha_0=0$, $\alpha=1$, and $\sigma_z^2=0.1$. The numbers of robot particles and source particles are $M_r=30$, $M_s=30$ for $K=1$, $M_r=100$, $M_s=100$ for $K=2$, and $M_r=300$, $M_s=300$ for $K=3$. We normalize the control input such that $\lVert\mathbf{c}_{n,k}\rVert_2=1$. In other words, each robot moves 1 meter in each control cycle. The variance of the control error is chosen such that $E\{\lVert\mathbf{n}_{n,k}\rVert^2_2\}=0.05$. To avoid collisions, the minimum tolerated distance between any two robots is $d_{\min}=4$. As in \cite{leitinger2019belief}, we introduce a small noise with variance $\sigma_s^2=0.1$ in the transition model of the source position for the sake of numerical stability, although the source remains stationary during the whole seeking process. When a robot is within $5$ meters of the source, it will stop moving but continue to measure distances to assist the other robots in approaching the source. The algorithm is terminated when either the number of control cycles reaches $500$ or all the robots are within $5$ meters of the source.
Except for the robot trajectories, all the other curves in the following figures depict the average results from 100 trials.

\begin{figure*}
	\normalsize
	\centering
	\subfigure[]{
		\includegraphics[width=1.5in]{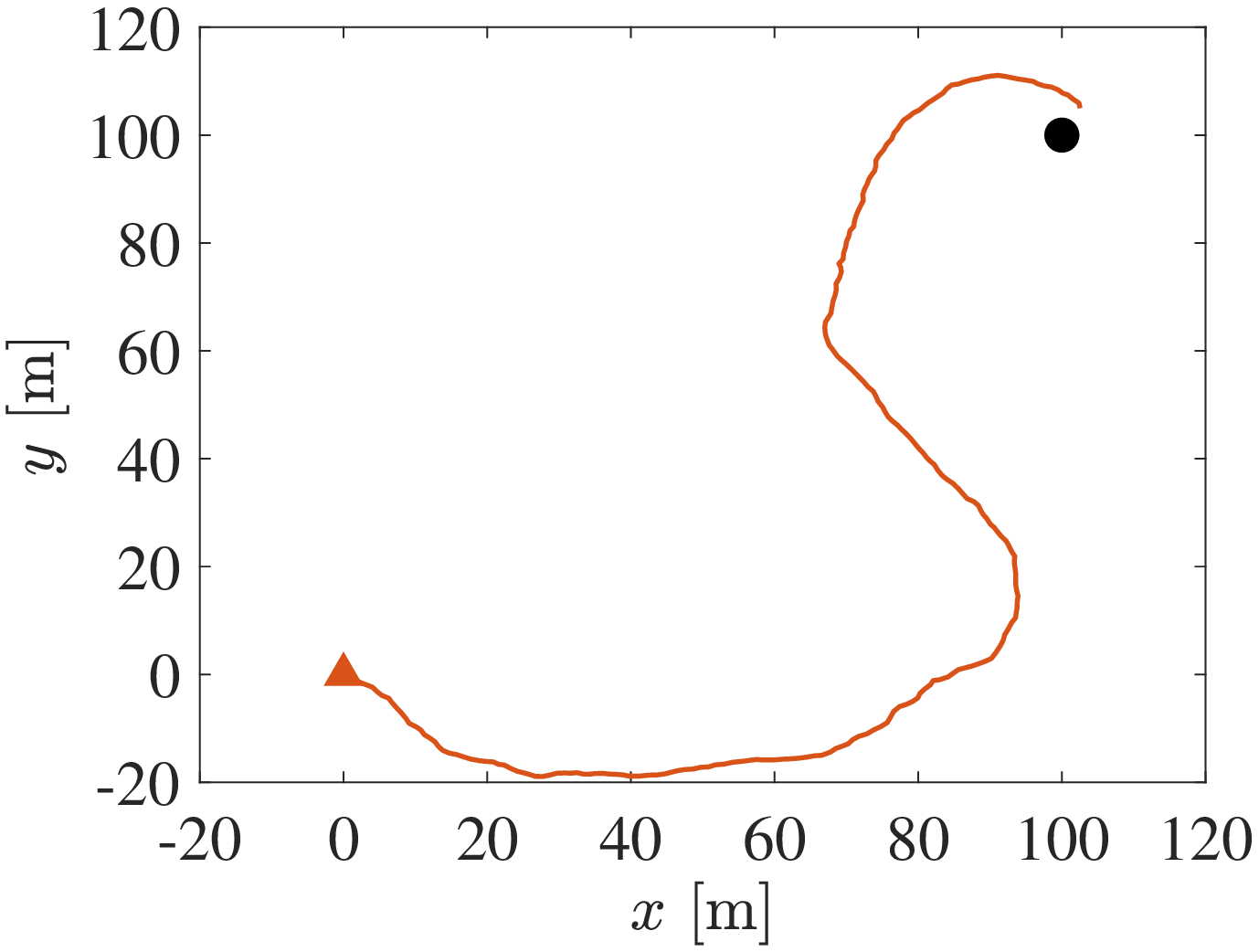}
	}
	\subfigure[]{
		\includegraphics[width=1.5in]{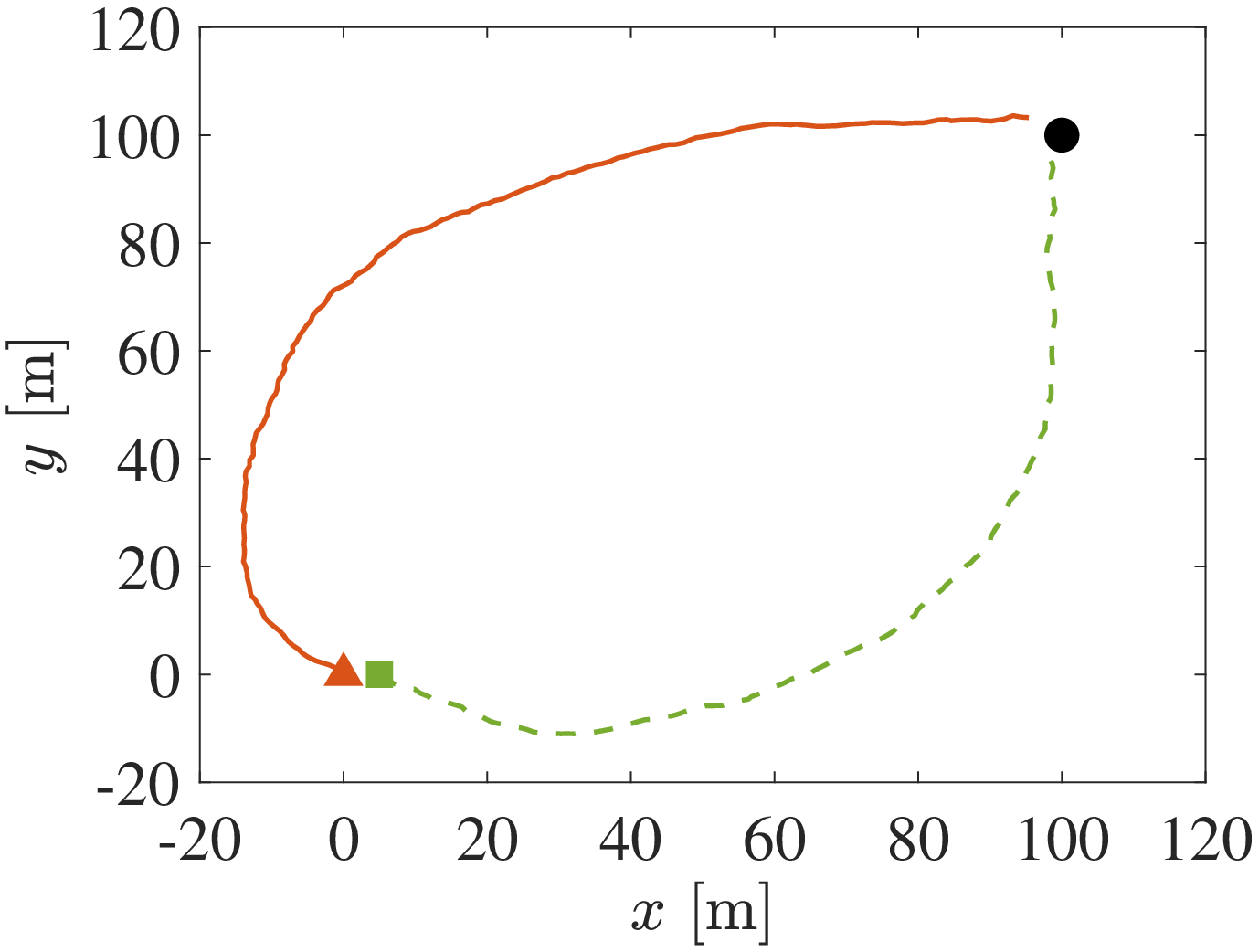}
	}
	\subfigure[\label{robot3}]{
		\includegraphics[width=1.5in]{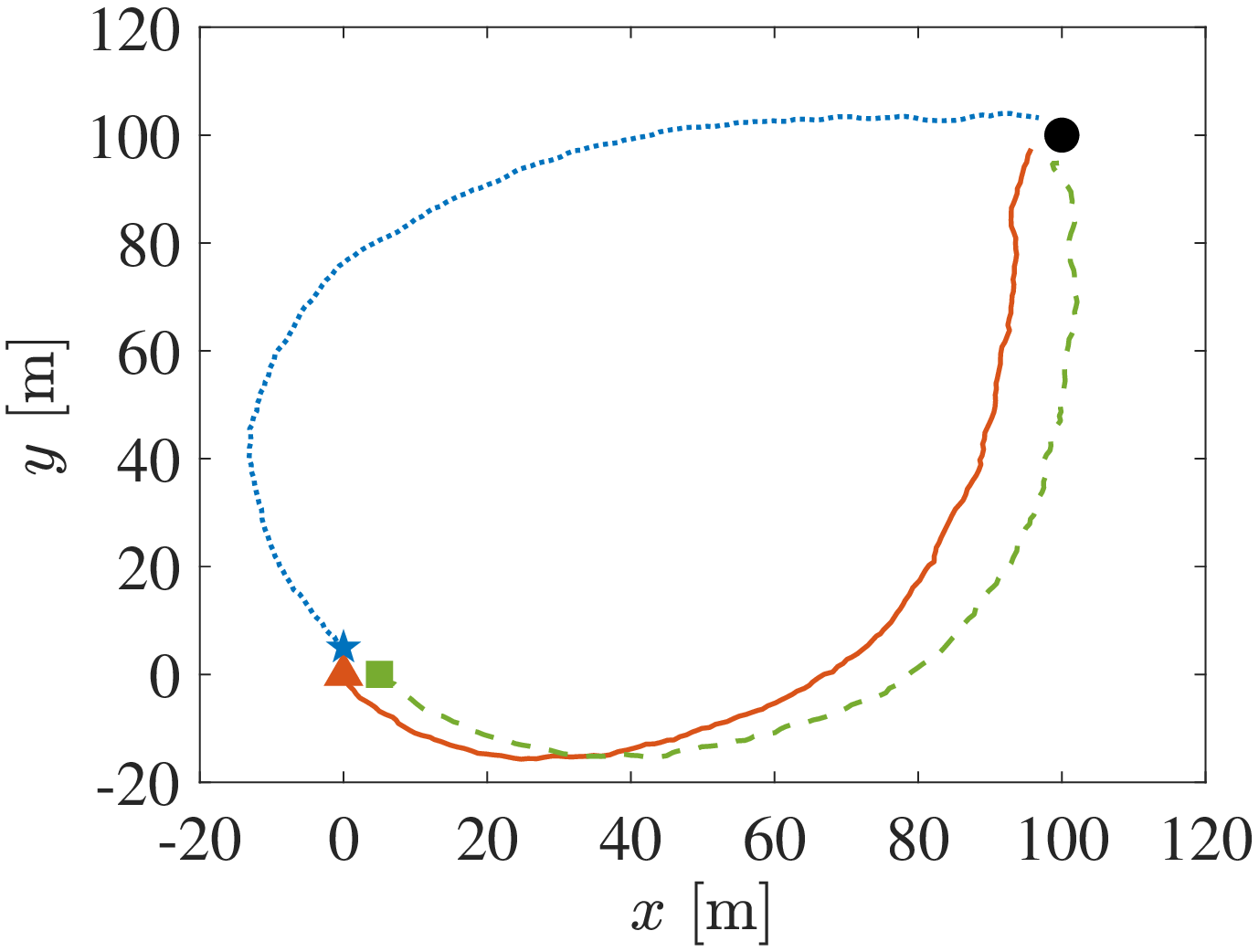}
	}
	\subfigure{
		\includegraphics[width=1.4in]{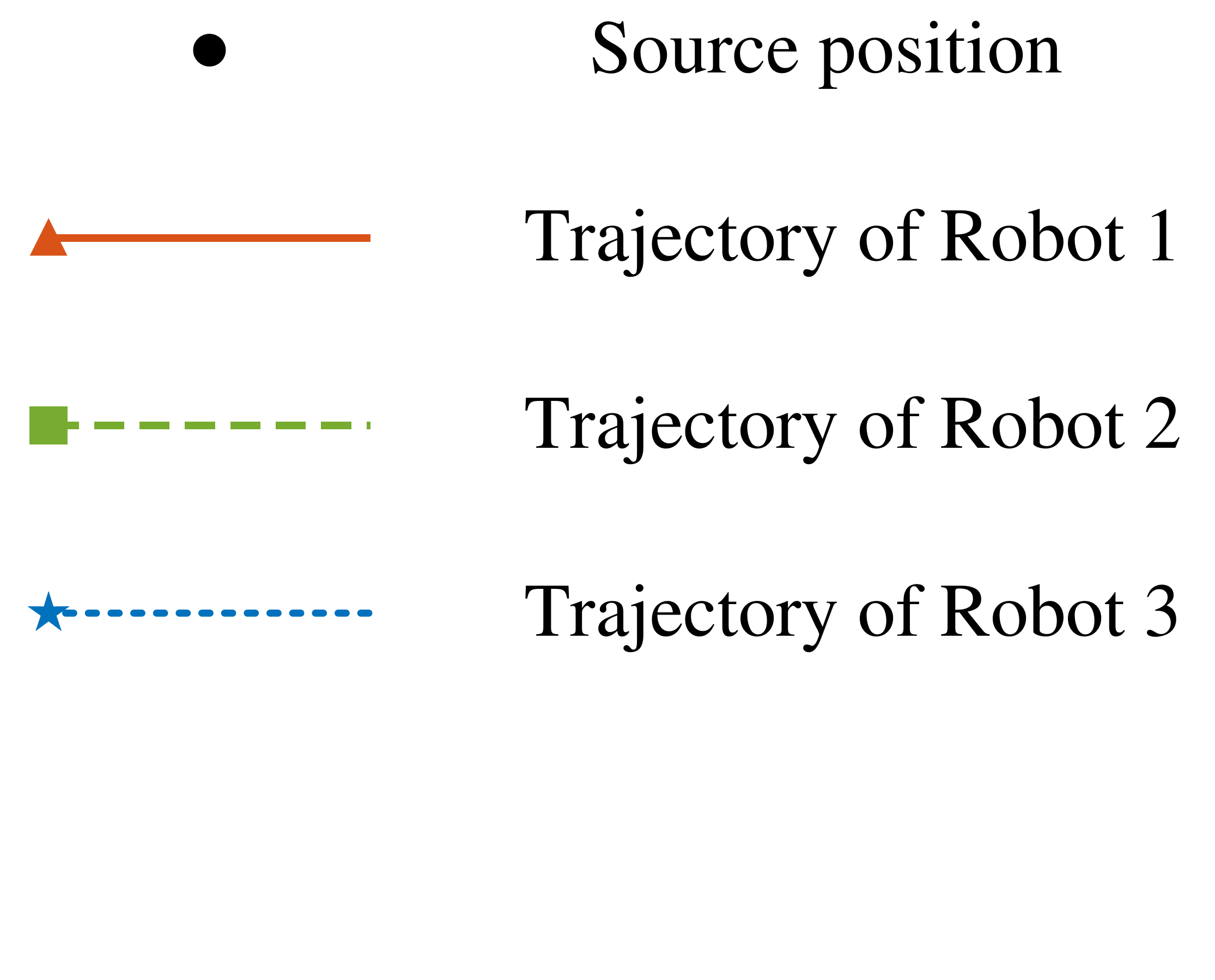}
	}
	\caption{Robot trajectories of the proposed source-seeking scheme with $K$ varying from $1$ to $3$.}
	\label{proposed trajectory}
	\hrulefill
	\vspace{-1.5em}
\end{figure*}

Fig. \ref{proposed trajectory} shows examples of the robot trajectories with the number of robots $K$ varying from $1$ to $3$, in which the triangle/square/star indicates the starting position of a robot. It can be observed that the trajectories are not straight lines from the starting positions to the source. In the cases of $K=2$ and $K=3$, the robots first spread out and then converge to the source. This is because there is a high uncertainty about the source position at the early stage, and thus moving directly toward the estimated source position possibly leads to a deviation from the true position. Spreading allows the robots to receive diverse distance measurements between the source and the robots. Hence, position information can be observed from different views. After the source position uncertainty reduces to a low level, the robots can gradually approach the source. Even when $K =1$, a twisty trajectory is observed, which is also for reducing the source position uncertainty. All these robot behaviors directly result from the information-theoretic control policy proposed in Section \ref{inforcontrol}. Note that robot 1 and robot 2 do not pass through the intersection point in Fig. \ref{robot3} in the same control cycle. In other words, the collision avoidance constraints are not violated.

\begin{figure}
	\centering
	\setlength{\abovecaptionskip}{0.2em}
	\subfigure[\label{aaaaa}]{
		\includegraphics[width=1.6in]{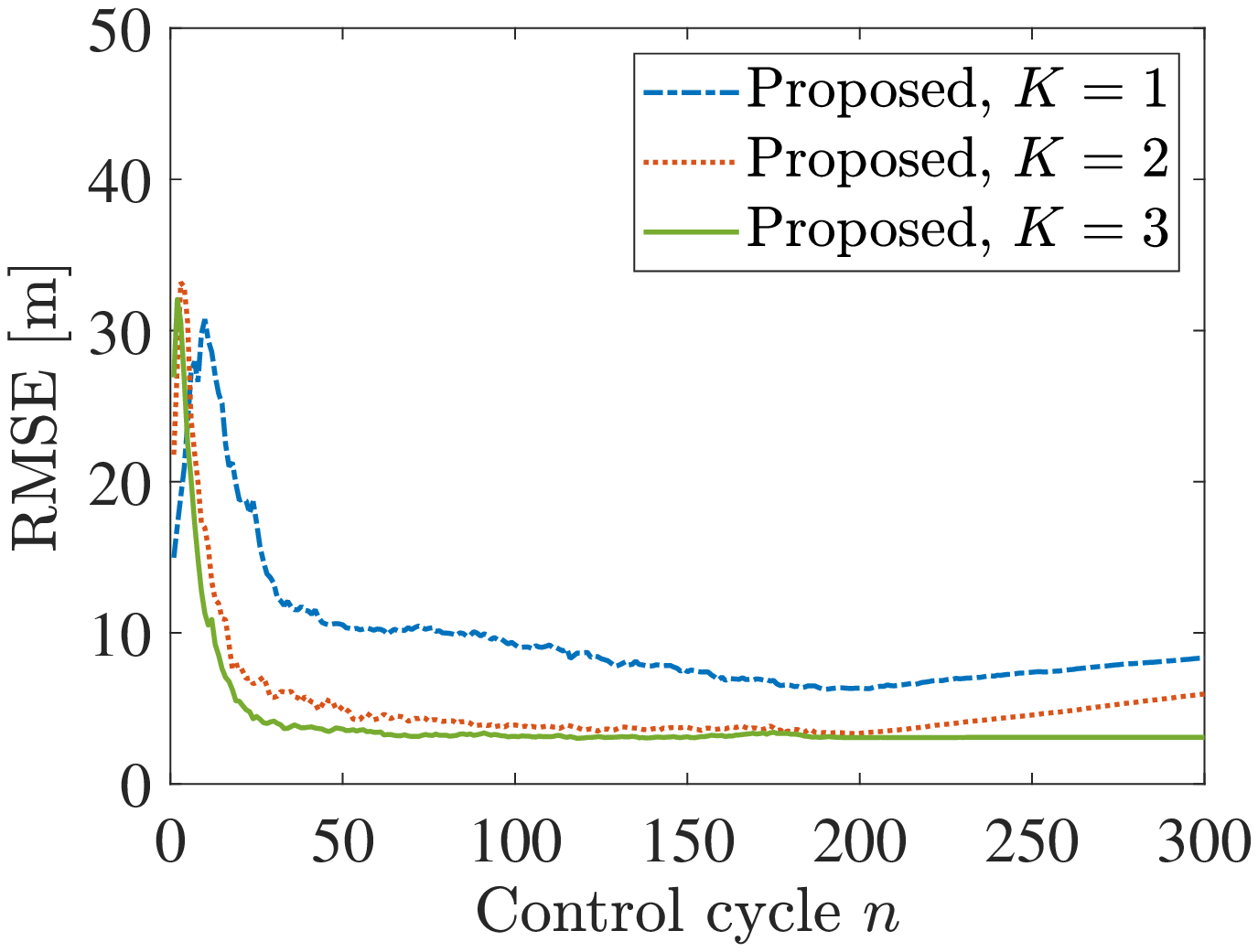}
	}
	\subfigure[\label{bbbbb}]{
		\includegraphics[width=1.6in]{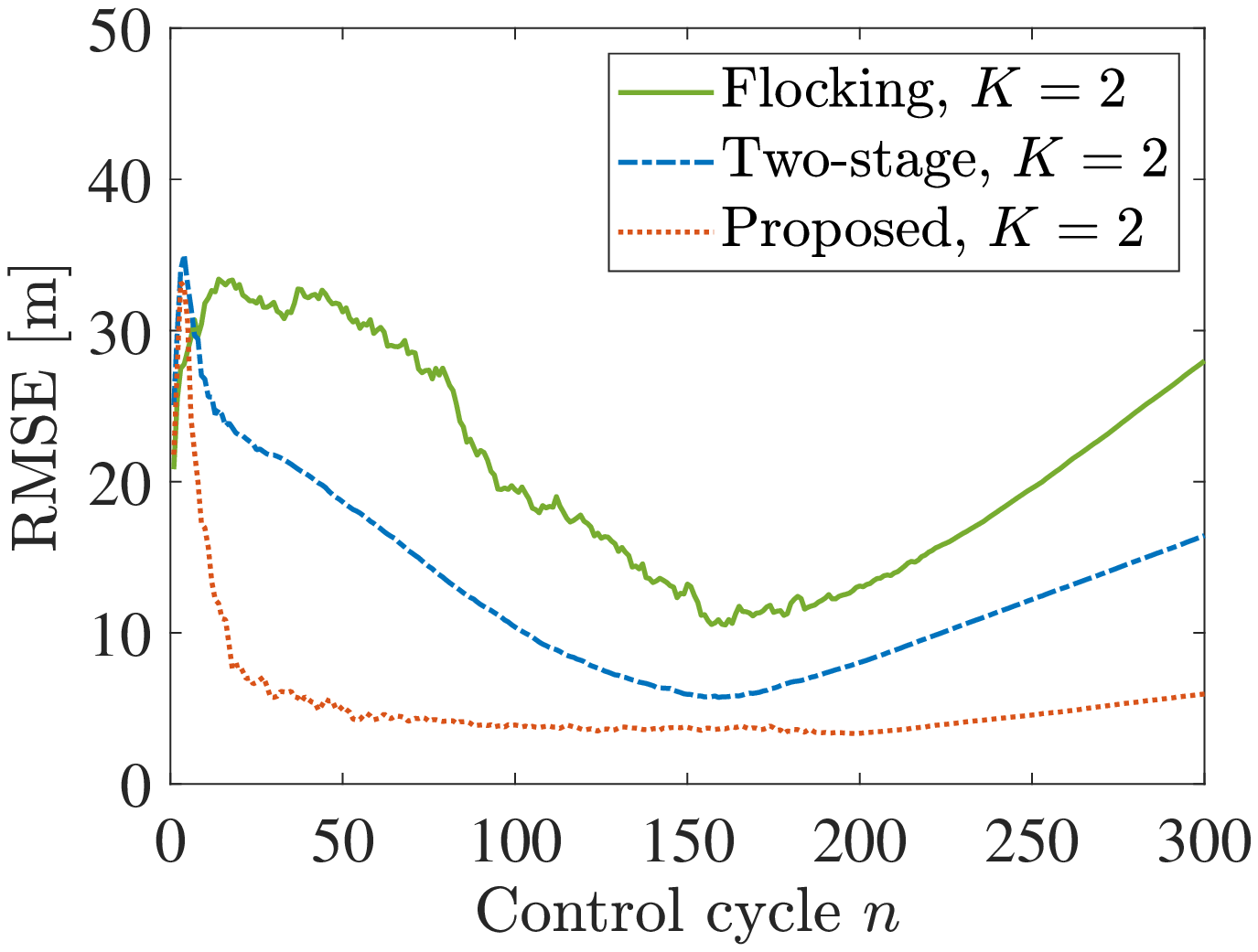}
	}
	\quad
	\subfigure[\label{ccccc}]{
		\includegraphics[width=1.6in]{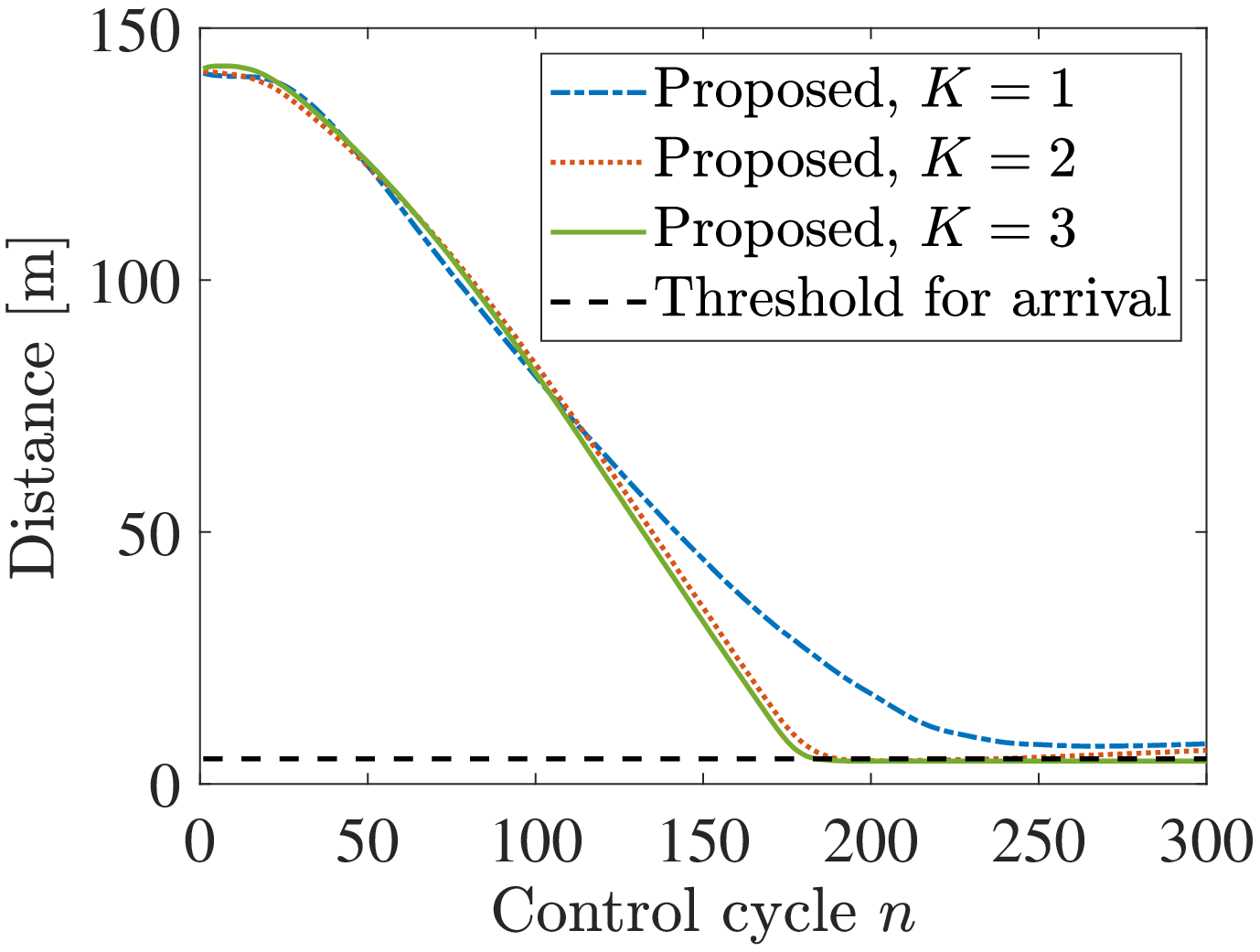}
	}
	\subfigure[\label{ddddd}]{
		\includegraphics[width=1.6in]{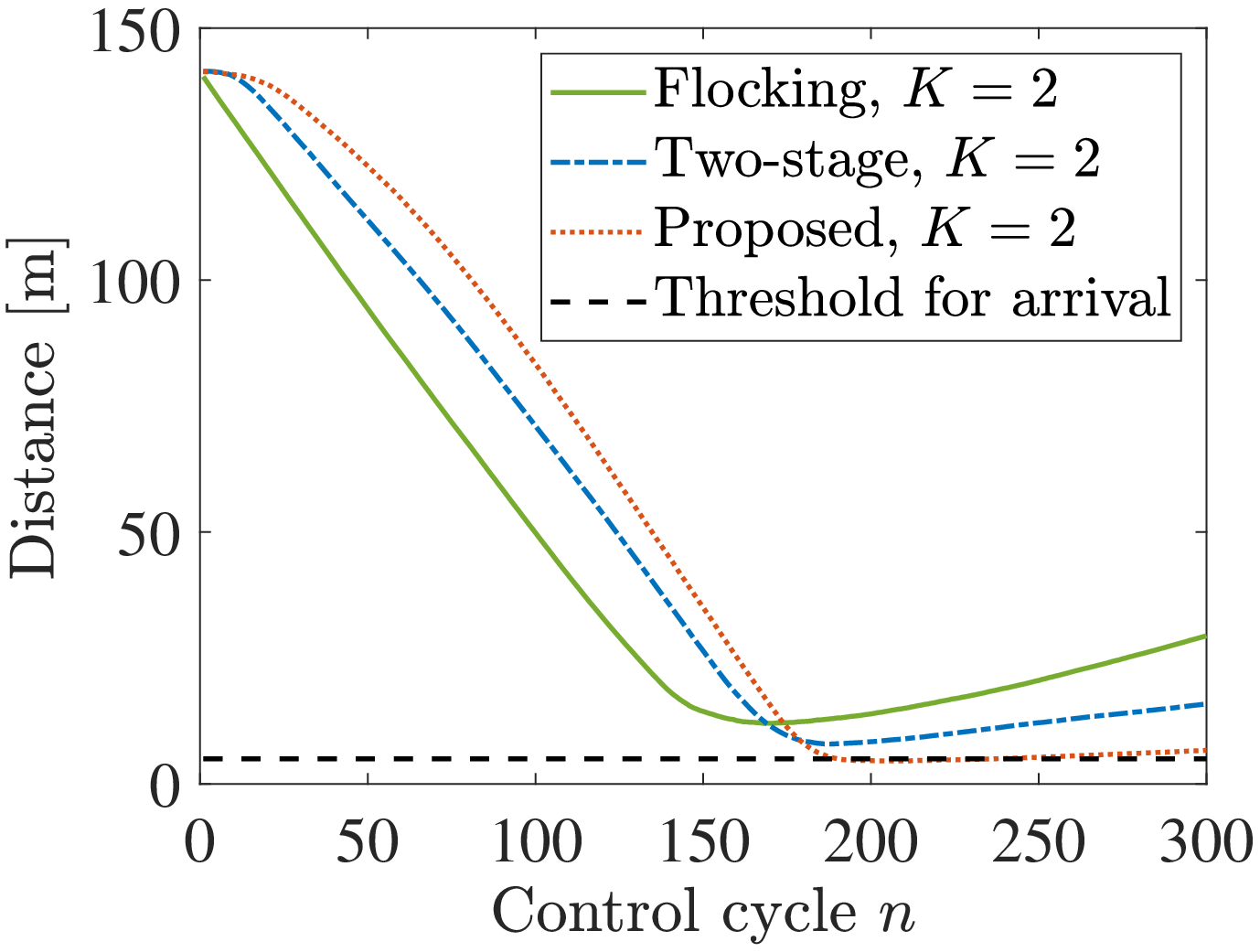}
	}
	\caption{(a) RMSE comparison with $K$ varying from 1 to 3. (b) RMSE comparison with different source-seeking schemes. (c) Comparison of the distances between robot 1 and the source with $K$ varying from 1 to 3. (d) Comparison of the distances between robot 1 and the source with different source-seeking schemes.}
	\label{MSE}
	\vspace{-1.5em}
\end{figure}

Fig. \ref{aaaaa} shows the root mean square error (RMSE) of the estimated source position with the number of robots $K$ varying from 1 to 3. In each trial, if all robots reach the source before $n=500$, the subsequent RMSE values are set to equal the one computed at the final positions of the robots. From the figure, a peak can be observed in each curve when $n$ is small, implying that there is a large localization uncertainty at the early stage. In addition, the peak becomes narrow as the number of robots increases. This is reasonable because the cooperation among the robots improves the localization efficiency and thus the source position uncertainty rapidly reduces as $n$ increases. It is noted that the RMSE does not decrease monotonically with $n$, since, in some trials, the estimated source position deviates from the true position due to the accumulated localization error during the source-seeking process. As a result, the robots fail to reach the source and gradually move away from the source, leading to large RMSE. For the other trials, the RMSE remains unchanged after the robots reach the source. Therefore, the overall RMSE increases when $n$ is large. 

In Fig. \ref{bbbbb}, we fix $K=2$ and compare the RMSE of the proposed source-seeking scheme with two benchmark schemes: a flocking scheme and a two-stage scheme (i.e., the trivial scheme discussed earlier). In the flocking scheme, the robots move directly toward the estimated source position in each control cycle. For fairness, we use the same Rao-Blackwellized particle filtering method for the cooperative localization. The only difference is that the objective is to maximize 
$	\sum_{k=1}^K\Vert\mathbf{x}_{N+1,k}-\mathbf{x}_{N+1,0}\rVert_2$. Compared with the source-to-robot distances, the robots are close to each other initially. Hence, the flocking scheme will generate similar control inputs for all the robots. Instead of spreading out, the robots will approach the estimated source position as a flock, and no measurement diversity can be used to reduce the large localization uncertainty in the early stage. As a result, the flocking scheme has the worst RMSE performance and the robots are most likely to deviate from the true position of the source and gradually move away.
In the two-stage scheme, we first use a conventional particle filtering algorithm to estimate the robot positions. Afterward, the estimated positions of the robots are regarded as the true positions to localize the source and design the control inputs. As we can see, the RMSE of the two-stage scheme is also higher than that of the proposed scheme because the proposed joint localization approach has leveraged the mutual benefits between the robot self-localization and the source localization and thus achieves better performance.

Fig. \ref{ccccc} shows the distances between one of the robot (we select $k=1$ without loss of generality) and the source with different $K$'s. We observe that the distances almost keep decreasing as $n$ increases. Further, with more robots cooperatively performing the task, the robots can move more efficiently to the source, which is particularly evident when $K$ is increased from 1 to 2. In Fig. \ref{ddddd}, we fix $K=2$ and compare the distances between the selected robot and the source in different schemes. In the two benchmarks, the robots can approach the source faster at the early stage. However, they are more likely to fail to reach the source eventually due to the large accumulated localization errors.

\section{Conclusions}

This paper investigated a simultaneous localization and source-seeking (SLASS) problem in multi-robot systems. A centralized algorithm has been proposed, including a cooperative localization algorithm to jointly estimate the positions of the source and robots, and an information-theoretic control policy for navigating the robots. Simulation results show that the proposed scheme outperforms two benchmarks, and the robot team can approach the source more efficiently as the number of robots increases. In future work, we will design decentralized and scalable SLASS algorithms that better suit our application scenario, considering the high computational complexity of the current algorithm.

\bibliography{document}
\bibliographystyle{IEEEtran}

\end{document}